\documentclass[12pt,a4paper]{article}
\usepackage{acromake}
\usepackage{cite}
\usepackage{topcapt}

\def\X{\hphantom{-}}
\def\NoCol{\multicolumn{1}{c}{}}
\def\mA{m_{\scriptscriptstyle A}}
\def\mB{m_{\scriptscriptstyle B}}
\def\LO{\Lambda^0}
\def\SO{\Sigma^0}
\def\SM{\Sigma^-}
\def\SP{\Sigma^+}

\def\XO{\Xi^0}
\def\XM{\Xi^-}
\def\Kl3{K_{\ell3}}
\def\vev#1{\left\langle#1\right\rangle}
\def\adhoc{{\em ad hoc\/}\xspace}
\def\eg{{\em e.g.}}
\def\etal{{\em et al.}}
\def\ie{{\em i.e.}}
\def\ft{{\it ft\/}\xspace}

\acromake{HSD}{HSD}{hyperon semi-leptonic decay}
\acromake{DHK}{DHK}{Donoghue, Holstein and Klimt}
\acromake{CoM}{CoM}{centre-of-mass}
\acromake{CKM}{CKM}{Cabibbo-Kobayashi-Maskawa}
\begin{document}
\title{
 \hfill\raisebox{2cm}[0pt][0pt]{\normalsize EPTCO-98-002}\\
 \bf SU(3) Breaking \\
 \bf in Hyperon Beta Decays: \\
 \bf a Prediction for $\mathbf{\XO\to\SP{e}\bar\nu}$ \\[6pt]
 \normalsize \rm [submitted to {\em Phys.~Rev.~Lett.}]
}
\author{
 Philip G. Ratcliffe\,\thanks{E-mail: pgr@fis.unico.it} \\
 \small \em Istituto di Scienze, Universit\`a di Milano---sede di Como,
 \\[-12pt]
 \small \em via Lucini 3, 22100 Como, Italy
 \\[-6pt]
 \small \rm and
 \\[-6pt]
 \small \em Istituto Nazionale di Fisica Nucleare---sezione di Milano
}
\date{June 1998}
\maketitle
\begin{abstract}
 On the basis of a previous analysis of hyperon semi-leptonic decay data, a
 prediction is presented for $g_1/f_1$ in the $\XO\to\SP{e}\bar\nu$
 $\beta$-decay. The analysis takes into account SU(3) breaking in this sector
 via the inclusion of mass-driven corrections. A rather precise measurement
 of the above channel by the KTeV experiment at Fermilab will shortly be
 available. Since the dependence on the SU(3) parameters, $F$ and $D$, is
 identical to that of the neutron $\beta$-decay, such a measurement will
 provide a rather stringent test of SU(3) and the models used to describe its
 violation in these decays. The prediction given here for the above decay is
 $g_1/f_1=1.17$, which leads to a rate of $0.80\times10^{6}$\,s$^{-1}$ and
 thus a branching fraction of $2.3\times10^{-4}$.
 \\[6pt]
 \\[6pt]
\end{abstract}
\newpage
\section{Introduction}

In recent years the precision of experimental \HSD data has improved steadily
\cite{Bourquin:1982ba,Hsueh:1988ar,Dwo90a,PDG96a} with parameters and rates
now known to within a few percent. Indeed, the present accuracy demands an
approach for applying corrections due to the breaking of SU(3). However,
there are several methods proposed in the literature; all describe the data
with varying degrees of success, from different starting points and with
differing output values for the parameters involved (\eg, $F$ and $D$).

The imminent release of an entirely new branching ratio, that of
$\XO\to\SP{e}\bar\nu$, will permit testing the various approaches. This
channel is being studied by the KTeV experiment at Fermilab\cite{Cheu:1997a}.
It is important to note that the angular correlations will also be measured
\cite{Swallow:pc}. It turns out that this particular channel has an axial
decay constant given by $g_1/f_1=F+D$, which in the absence of SU(3) breaking
would therefore be identical to that of neutron $\beta$-decay (by far the
most precisely known). Thus, provided both $\Gamma$ and $g_1/f_1$ are
measured independently, such a channel can provide a sensitive test of the
methods used to describe SU(3) symmetry breaking in this sector.

\section{Hyperon Semi-Leptonic Decay Data}

The present situation with regard to the \HSD data is shown in
table~\ref{tab:data}, which represents the useful available knowledge.
\begin{table}[hbt]
\begin{center}
\topcaption{\label{tab:data}
 The hyperon semi-leptonic data used in this analysis~\protect\cite{PDG96a},
 $g_1/f_1$ indicates the value as extracted from angular correlations.
 The last column shows the SU(3) formula for $g_1/f_1$.}
\begin{minipage}{\textwidth}\let\footnoterule\relax
\begin{center}
$\begin{array}{|l@{\,\to\,}l|l@{\,\pm\,}l|l@{\,\pm\,}l|l@{\,\pm\,}l|l|}
 \hline
   \multicolumn{1}{|c }{} &
 & \multicolumn{4}{ c|}{\Gamma\quad(10^6\,\mbox{s}^{-1})}
 & \multicolumn{3}{ c|}{g_1/f_1}
 \\ \cline{3-9}
   \multicolumn{2}{|c|}{\raisebox{2ex}[0pt][0pt]{Decay}}
 & \multicolumn{2}{ c|}{\ell=e^-}
 & \multicolumn{2}{ c|}{\ell=\mu^-}
 & \multicolumn{2}{ c|}{\ell=e^-}
 & \multicolumn{1}{ c|}{\mbox{SU(3)}}
 \\ \hline
 n   & p\,\ell\bar\nu & 1.1274 & 0.0025
 \,\footnote{The rate is given in $10^{-3}\,$s$^{-1}$.}& \NoCol &
 &\X1.2601 & 0.0025 & F+D
 \\ \hline
 \LO & p\,\ell\bar\nu & 3.161  & 0.058  & 0.60   & 0.13
 &\X0.718 & 0.015   & F+D/3
 \\ \hline
 \SM & n\,\ell\bar\nu & 6.88   & 0.23   & 3.04   & 0.27
 & -0.340 & 0.017   & F-D
 \\ \hline
 \SM & \LO\ell\bar\nu & 0.387  & 0.018  & \NoCol &
 & \NoCol &         & -\sqrt{\frac23}\,D
 \,\footnote{As $f_1=0$, the absolute expression for is $g_1$ given.}
 \\ \hline
 \SP & \LO\bar\ell\nu & 0.250  & 0.063  & \NoCol &
 & \NoCol &         & -\sqrt{\frac23}\,D\,^b
 \\ \hline
 \XM & \LO\ell\bar\nu & 3.35   & 0.37
 \,\footnote{A scale factor of 2 is included, following
 the PDG practice for discrepant data.} & 2.1    & 2.1
 \,\footnote{These data are not used in the fits.}
 &\X0.25  & 0.05    & F-D/3
 \\ \hline
 \XM & \SO\ell\bar\nu & 0.53   & 0.10   & \NoCol &
 & \NoCol &         & F+D
 \\ \hline
\end{array}$
\end{center}
\end{minipage}
\end{center}
\end{table}
As discussed in \cite{Ratcliffe:1996fk}, the disagreement between the neutron
lifetime and the value of $g_1/f_1$ extracted from $\beta$-decay angular
correlations \cite{Towner:1995za} requires some care, to avoid clouding the
issue of SU(3) breaking. The present value of the neutron lifetime is
$887.0\pm2.0\,$s and $g_1/f_1$ (from angular correlations) is
1.2601(25)\,\footnote{The slight change in the value since the publication of
\protect\cite{Ratcliffe:1996fk} has no visible effect on any of the fits.}
\cite{PDG96a}; \ie, both are known independently to approximately 0.2\%. The
value of the relevant \CKM matrix element extracted from the \ft values of
the eight super-allowed nuclear $\beta$-decay Fermi transitions is
$V_{ud}=0.9740(5)$ \cite{Towner:1995kk}. This is to be compared with the
values: $V_{ud}=0.9795(20)$, from the neutron lifetime and $g_1/f_1$, and
$V_{ud}=0.9758(4)$, from the so-called $\Kl3$ decays ($V_{us}=0.2188(16)$
\cite{Garcia:1992pu}).

The displacements from the central values are all very small, $<0.2\%$. Thus,
to neutralise the contribution of the neutron discrepancy to the global
$\chi^2$, a mean value for $V_{ud}$ is first extracted from the nuclear \ft
and $\Kl3$ data. Then using this value, a combined fit to the
$\Gamma_{n\to{p}}$ and $g_1/f_1$ is made. Finally, the errors of the
$\Gamma_{n\to{p}}$, $g_1/f_1$ and mean $V_{ud}$ values are multiplied by the
$\sqrt{\chi^2}$ so obtained; these are used in all fits:
\begin{eqnarray}
 \Gamma(n\to p\ell\bar\nu) &=& (1.1274 \pm 0.0055) \times 10^{-3}s^{-1} \\
 g_1/f_1                   &=& \hphantom{(} 1.2601 \pm 0.0055           \\
 V_{ud}                    &=& \hphantom{(} 0.9752 \pm 0.0007.
\end{eqnarray}

\section{SU(3) Analyses \label{sec:analyses}}

In table \ref{tab:su3fits} the results of a series of fits to the \HSD data
are displayed: the symmetric fit uses three parameters ($F$, $D$ and
$V_{ud}$), and the SU(3) breaking is described by one further parameter
(described in the following). We use the mean value obtained from the
combined nuclear \ft analysis and $\Kl3$ decays just described, and impose
the unitarity constraint $V_{ud}^2+V_{us}^2=1$ (neglecting
$V_{ub}=0.0033\pm0.0008$ \cite{PDG96a}).
\begin{table}[hbt]
\begin{center}
\topcaption{\label{tab:su3fits}
 SU(3) symmetric and breaking fits to the modified data, including the
 external $V_{ud}$ from nuclear \ft and $\Kl3$ analyses (see the text for
 details).}
$\begin{array}{|c|l@{\,\pm\,}l|l@{\,\pm\,}l|l@{\,\pm\,}l|c|c|}
 \hline
 & \multicolumn{6}{ c|}{\mbox{Parameters}} & &
 \\ \cline{2-7}
   \multicolumn{1}{|c|}{\raisebox{2ex}[0pt][0pt]{Fit}}
 & \multicolumn{2}{ c|}{V_{ud}}
 & \multicolumn{2}{ c|}{F}
 & \multicolumn{2}{ c|}{D}
 & \multicolumn{1}{ c|}{\raisebox{2ex}[0pt][0pt]{$\chi^2\!/$DoF}}
 & \multicolumn{1}{ c|}{\raisebox{2ex}[0pt][0pt]{$F/D$}}
 \\ \hline
 \mbox{Sym.} & 0.9749 & 0.0003 & 0.465 & 0.006 & 0.798 & 0.006 & 2.3 & 0.582
 \\ \hline
 \mbox{A}    & 0.9743 & 0.0004 & 0.460 & 0.006 & 0.806 & 0.006 & 1.2 & 0.571
 \\ \hline
 \mbox{B}    & 0.9744 & 0.0004 & 0.459 & 0.006 & 0.807 & 0.006 & 1.2 & 0.571
 \\ \hline
\end{array}$
\end{center}
\end{table}
The parametrisations of the SU(3) breaking used are the so-called \CoM
correction \cite{Bog68a} (fit A), which is described in detail in
\cite{Ratcliffe:1996fk}, and an alternative breaking scheme (fit B), using an
SU(3) motivated mass dependence for the axial couplings
\cite{Ratcliffe:1997ys}.

Approach A is to apply \CoM or recoil corrections to the axial coupling
constant for the process $A\to{B}\ell\nu$ according to the following formula
\cite{Don87a}:
\begin{equation}
 g_1 = g_1^{\mathrm{SU(3)}} \; \left\{ 1 - \frac{\vev{p^2}}{3\mA\mB} \,
 \left[ \frac14 + \frac{3\mB}{8\mA} + \frac{3\mA}{8\mB} \right] \right\}.
\label{eq:CoM}
\end{equation}
A similar correction to the vector piece is entirely negligible (in
accordance with the Ademollo-Gatto theorem \cite{Ade64a}) and thus here $f_1$
is taken to have its na\"{\i}ve SU(3) CVC value. The mean momentum squared,
$\vev{p^2}$, is calculated by \DHK using a bag model to be
$0.43$\,GeV$^2$, 
here it is left as a free parameter and is determined in fit A to be
$0.43\pm0.11$\,GeV$^2$. The results for approach B are necessarily rather
similar as it effectively corresponds to a linearisation of eq.~\ref{eq:CoM}.

As can be seen from table~\ref{tab:su3fits}, the data clearly indicate the
presence of SU(3) breaking, which is well described by the correction schemes
adopted. Note also that the value of the ratio $F/D$ is largely unaffected by
the breaking, changing by less than 2\%, and should thus not be considered as
an indicator of the importance of SU(3) breaking. In both schemes a possible
additional breaking in the $|\Delta{S}=1|$ decays has been neglected; in
previous fits this was found to be at most about 2\%; in any case, it is
essentially absorbed into the extracted value of $\sin^2\theta_C$ and has
negligible effect on $F$ and $D$.

\section{The prediction for $\mathbf{\XO\to\SP{e}\bar\nu}$}

At this point a prediction is possible for any of the remaining HSD's: in
particular, the $\XO\to\SP{e}\bar\nu$ $\beta$-decay. Since independent
measurements of both the decay rate and the angular correlations should be
obtained, this measurement will, in principle, simultaneously allow separate
control over the smallness of the corrections associated with the
$|\Delta{S}=1|$ decays (assumed here) and of the validity of corrections
applied in the above analysis.

The values obtained are shown in table~\ref{tab:prediction}.
\begin{table}[hbt]
\begin{center}
\topcaption{\label{tab:prediction}
 The values obtained for the axial coupling ($g_1/f_1$), rate ($\Gamma$) and
 branching fraction ($B$) for the $\XO\to\SP{e}\bar\nu$ $\beta$-decay. The
 errors quoted are purely those returned by the fitting routine.}
\begin{minipage}{\textwidth}\let\footnoterule\relax
\begin{center}
$\begin{array}{|c|l@{\;\pm\;}l|l@{\;\pm\;}l|l@{\;\pm\;}l|}
 \hline
   Fit
 & \multicolumn{2}{ c|}{g_1/f_1}
 & \multicolumn{2}{ c|}{\Gamma\;(10^6\,\mbox{s}^{-1})}
 & \multicolumn{2}{ c|}{B\;(10^{-4})}
 \\ \hline
 \mbox{Sym.} & \multicolumn{2}{l|}{1.26\,\footnote{No serious error can be
 associated with the value of $g_1/f_1$ in the symmetric fit as it should be
 precisely that of the neutron $\beta$-decay.}}
                           & 0.89 & 0.01 & 2.58 & 0.05
 \\ \hline
 \mbox{A}    & 1.17 & 0.03 & 0.80 & 0.03 & 2.32 & 0.10
 \\ \hline
 \mbox{B}    & 1.14 & 0.03 & 0.78 & 0.03 & 2.26 & 0.12
 \\ \hline
\end{array}$
\end{center}
\end{minipage}
\end{center}
\end{table}
Included in the error for the branching fraction is the contribution from the
error on the total decay width of the $\XO$, which is about 3\%, the others
are those returned by the global fit. The difference between the two breaking
fits, A and B, is an indication of the expected systematic uncertainty arising
from this type of description, and which we thus estimate to be less than 3\%.
The spread is also small compared to the shift from the na\"{\i}ve values.

\section{Conclusions \label{sec:conc}}

First of all, as has been demonstrated in detail elsewhere
\cite{Ratcliffe:1997ys}, the axial couplings extracted from the hyperon
decays are well described by a parametrisation motivated by the mass
differences in the baryon octet. The results discussed above permit a precise
prediction for the $\XO\to\SP{e}\bar\nu$ $\beta$-decay: here both
the $g_1/f_1$ and the expected decay rate have been presented. The values
given may be compared to another prediction in the literature due to
Flores-Mendieta, Jenkins and Manohar \cite{Flores-Mendieta:1998ii}. In a
breaking scheme based on the $1/N_c$ expansion, the authors cited find a
value for $g_1/f_1$ considerably smaller than that predicted here: $f_1=1.12$
and $g_1=1.02$, or $g_1/f_1=0.91$ (their fit B), which leads to a rate of
$0.68\times10^6\,$s$^{-1}$. Their prediction for the SU(3) parameters is
$F/D=0.46$, to be compared with $0.57$ above. In an alternative fit, where
$f_1$ is left at its SU(3) value, they obtain $g_1/f_1=1.03$ (their fit A)
and $0.65\times10^6\,$s$^{-1}$. In either case both $g_1/f_1$ and the rate
are considerably smaller than the results of the present analysis, which in
turn are considerably smaller than a na\"{\i}ve fit. Thus, the various
possibilities should be distinguishable in an experiment with good
statistics, such as KTeV.

It should perhaps be mentioned that the Flores-Mendieta \etal\ fit also
includes data on the baryon decuplet non-leptonic decays, which in fact
dominate the final results. The overall fit, according to the value of
$\chi^2$ returned, is rather poor. Moreover, their approach applied to the
\HSD data alone produces results similar to those reported in this paper
\cite{Manohar:pc}.

Secondly, in this analysis, as too in \cite{Flores-Mendieta:1998ii}, the
possibility of a weak electric ($g_2$) contribution has been neglected. It is
therefore worth remarking that experimental data on the $\SM\to{n}e\bar\nu$
$\beta$-decay \cite{Hsueh:1988ar} indicate that such a second-class current
contribution may be non-negligible. Indeed, the data marginally prefer a
sizable $g_2$ and thus a much reduced value for $g_1$ there. If such were the
case, then the question would also arise as to the relevance of $g_2$ in
other decays, where the experimental analysis has typically assumed it zero.

Thirdly, a paper often quoted in the literature as providing evidence for
large breaking effects, similar to those found in
\cite{Flores-Mendieta:1998ii}, is that of Ehrnsperger and Sch\"afer
\cite{Ehr94a}. There the authors apply an \adhoc one-parameter ($a$ below)
correction to the angular correlation data alone:
\begin{equation}
 F/D = (F/D)^{\mathrm{SU(3)}} \;
 \left[ 1 + a\frac{(\mA+\mB)-(m_n+m_p)}{(\mA+\mB)+(m_n+m_p)} \right],
\end{equation}
where $a$ is found to be $\sim2.7$; thus, the limiting value of $F/D$ is
$0.49\pm0.08$ (note the large error). However, since the breaking is treated as
affecting only the ratio $F/D$ and not the sum, such a solution implies that
the $\XO\to\SP\ell\bar\nu$ decay has $g_1/f_1$ identical to that of the
neutron despite the enormous mass shift.

Finally, before closing, let us recall another decay for which the rates and
angular correlations are also expected to have very large corrections and
thus to be highly sensitive to SU(3) breaking: namely, $\XM\to\SO{e}\nu$.
Here too, the fact that $g_1/f_1=F+D$ makes it highly desirable to improve on
the present limited experimental knowledge for this process.

\section{Acknowledgments}

The author is most grateful to Prof.~E.C. Swallow for much helpful
information and comment.

\end{document}